\documentclass[prl,10pt,twocolumn,showpacs]{revtex4}

\usepackage{graphicx}
\usepackage{color}
\usepackage{epsfig}
\usepackage{amsmath}
\usepackage{amstext}
\usepackage{amssymb}

\usepackage[running]{lineno}
\DeclareGraphicsExtensions{.eps,.ps}

\begin{document}

\author{Mateusz Cholascinski}

\author{Ravindra W. Chhajlany}

\affiliation{Institute of Physics, Adam Mickiewicz University, 61-614
Pozna\'n, Poland}

\date{\today}

\title{ Stabilized parametric Cooper-pair pumping in a
           linear array of coupled Josephson junctions}

\begin{abstract}
  We present an experimentally realizable stabilized charge pumping
  scheme in a linear array of Cooper-pair boxes.  The system design
  intrinsically protects the pumping mechanism from severe errors,
  especially current reversal and spontaneous charge excitation. The
  quantum Zeno effect is implemented to further diminish pumping
  errors. The characteristics of this scheme are considered from the
  perspective of improving the current standard. Such an improvement
  bears relevence to the closure of the so-called measurement triangle 
%   which
%   enables the relation of the current, voltage and frequency by
%  fundamental constants 
(see D. Averin [Nature 434, 285 (2005)]).
\end{abstract}

\pacs{74.81.Fa 03.65.Xp 73.23.-b 85.25.Cp}

\maketitle
% \tableofcontents

Adiabatic variation of parameters in a quantum system results in
evolution in which population of each of the energy levels is
conserved (adiabatic following). It is well known that, during such
evolution, the energy levels acquire two types of phases: time
dependent dynamical phases and time independent geometrical phases
determined only by the geometry of the path traversed in parameter
space.  The property of adiabatic following can be also used to induce
transport through a system, {\it viz.} a slowly transported potential
well carries its bounded particles.  \\ For closed paths, the initial
and final states (usually) differ only by a phase factor \cite{berry},
but the system evolution during the cyclic variation can still be
characterized by non-vanishing transport properties. This phenomenon,
known as parametric pumping, is observable in mezoscopic devices,
where during each pumping cycle a definite charge
% (number of electrons)
is transported through the system
\cite{pothier,pekola1,keller,pekola,niskanen,niskanen2}. Quantum pumps
are hoped to transform an AC signal of frequency $f$ into DC current
given by the relation $I=fQ$, where $Q=ne$ for normal devices, and
$Q=2ne$ for superconducting pumps ($n$ is the total charge transported
per cycle). If operated with perfect accuracy, quantum pumps could be
utilized to establish a standard of current. This is needed to close
the measurement triangle relating voltage, current and frequency by
fundamental constants \cite{zorin}.  However, nonadiabatic
corrections, uncontrolled tunneling, and sensitivity of the devices to
parametric fluctuations smear the output signal. An intristic
stabilizing property, like quantization of magnetic flux, crucial for
the voltage standard, is absent in the parametric charge pumps. Thus
the only way to stabilize the performance is to optimize the design of
such devices. E.g. in Ref.  \onlinecite{niskanen} enhanced control
over the tunelling amplitudes in the so-called Cooper pair ``sluice''
resulted in improved characteristics of the pumping scheme.

In this Letter, we combine advantages of the controllable Josephson
couplings used in the ``sluice'' and the array-like design of earlier
proposals \cite{pekola,toppari}. Our scheme is analyzed in the
framework of adiabatic passage in an effective two-level system.
Electrostatic coupling of the Josephson junctions forming the arrays
results in increased separation of the operational subspace from the
excited states. The array design also suppresses the most severe
process affecting the pumping accuracy -- current reversal. This
suppression increases exponentially with the number of junctions in
the system. Since after each pumping sequence the system is to be
found in a known {\em charge} state, we also make use of the quantum
Zeno effect to diminish the population of unwanted states.

The core mechanism of the adiabatic passage can be understood using
the simplest nontrivial -- two-level system. Suppose that we start at
degeneracy (zero field in the spin-$1/2$ language) and spin pointing
``up'' in the $z$ direction. Then the field is increased along the
spin direction and traverses a path in the $x-z$ plane which
approaches the degeneracy again from the negative $z$ direction. For
adiabatic pulses, the system remains in the ground state and follows
the field. However, at the end the spin points in the negative
$z$ direction. This simple example shows that although the system is
in a nondegenerate level during the passage and the traversed path in
the parameter space is closed the resulting transformation is highly
nontrivial as the spin direction is reversed. This is due to
the level crossing occurring at the beginning and  end of the path.
For zero field the states are degenerate, and this freedom of defining
the orthogonal basis results in discontinuity of the energy
eigenstates.

\begin{figure}[t] %% h=here, t=top,p=page of floats
\centerline{\resizebox{0.25\textheight}{!}{\rotatebox{0}
{\includegraphics{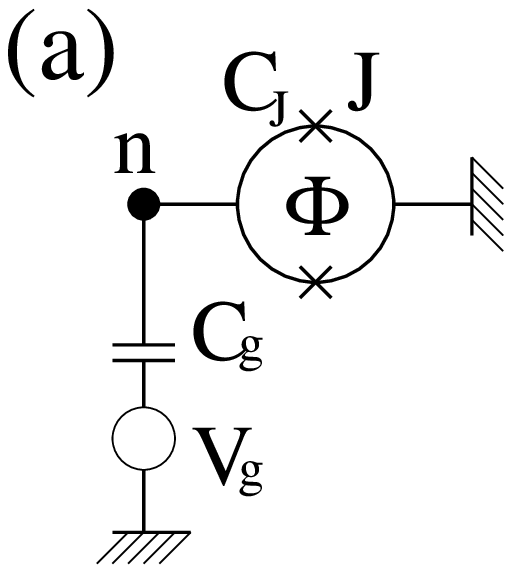}\hspace{1cm}\includegraphics{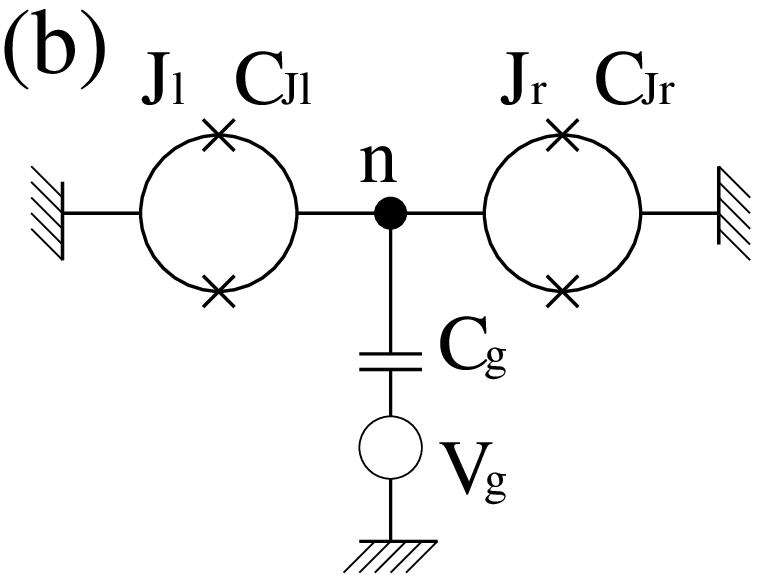}}}}
\caption{(a) Cooper pair box\cite{makhlin}. Electrostatic potential of
  the superconducting island (black node) is controlled by the gate
  voltage $V_g$. The Josephson tunelling amplitude $J$ is tuned by
  external flux $\Phi $. (b) A Cooper-pair box with two SQUIDS.}
\label{fi:basic}
\end{figure}

A Cooper-pair box with tunable Josephson coupling \cite{makhlin} [see
Fig.~\ref{fi:basic}(a)], described by the Hamiltonian $H= -1/2 B_z
\sigma _z - 1/2 B_x \sigma _x$, where $B_x \equiv J (\Phi ) = 2 J_0
\cos (\pi \Phi / \Phi_{0}) $, and $B_z = 4 E_C (1 - 2n_g)$, mimics a
spin-1/2 system. Here $J_0$ is the Josephson coupling of the junctions
in the SQUID (Superconducting Quantum Interference Device) coupling
the box to the reservoir, $\Phi $ and $\Phi_{0}$ the external applied
flux through the SQUID and flux quantum respectively, $E_C = e^2/ (2
C_g + 2 C_J)$ is the single-electron charging energy of the box and
$n_g = C_g V_g /2e$ is the dimensionless gate charge. 
%(or dimensionless gate voltage tuning the electrostatic potential of
%the box).
The two-level approximation holds if the system is operated in
the charge regime ($J_0 \ll E_C$) and $n_g$ is close to $1/2$. Then
the only relevant states are characterized by zero or one Cooper pair
in the box, corresponding to spin parallel and antiparallel to the
$z$ direction respectively.  Performing the procedure
described above on the system, we can thus transport a Cooper pair to
or from a reservoir into an empty box, depending on the sequence of
pulses.

\begin{figure}[t] %% h=here, t=top,p=page of floats
\centerline{\resizebox{0.3\textheight}{!}{\rotatebox{0}
{\includegraphics{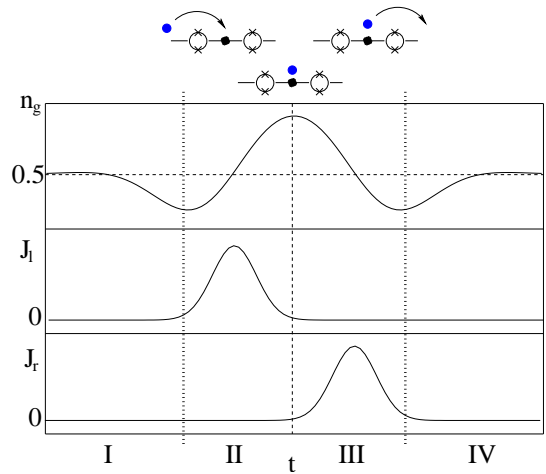}}}}
\caption{A pulse sequence applied to the simple pump that induces flow
of one Cooper pair through the system. The exact shape of, and
area under the pulses do not influence the process as long as the
adiabaticity condition holds.}
\label{fi:pulses}
\end{figure}

To induce transport through the Josephson-junction system we use two
SQUIDs which now serve as terminals [see Fig.\ref{fi:basic}(b)]. Using
one of the terminals to transport a Cooper pair into the box (stage
one) and another to transport the same pair to the reservoir (stage
two) we generate current $I = 2ef$. Exemplary pulses of a full cycle
are depicted in Fig.~\ref{fi:pulses}.  No transport takes place during
parts I and IV with the system in a definite (zero) charge state, so
we can limit the cycle to the sequences II and III only. This
simplification minimizes the cycle duration and eliminates
degeneracies from the procedure (this system is identical to the
Cooper pair ``sluice'' \cite{niskanen}).

Let us comment briefly on the possible sources of imperfections in the
pump performance. The nonadiabatic corrections leave the system in an
unknown superposition of charge states after the full cycles (instead
of a definite charge state). These corrections whilst usually small,
can accumulate over time. Fluctuations of external parameters affect
the scheme similarly, though fluctuations in external flux have
a much more severe effect than those in the gate voltage, especially
when the SQUIDs are not ideally symmetric. The terminals cannot then
be completely closed, due to residual Josephson coupling, and the
fluctuations induce transitions between charge states. The error
related to this feature is the hardest to eliminate -- since there is
nonzero tunneling to the reservoir via both terminals all the time.
Cooper pairs can accidentally be transported in the wrong direction,
leaving no trace of the error in cases when the the final charge state
is identical to the expected state. In the following we show that
the array design combined with Zeno projections  strongly
suppresses these pumping errors.

\begin{figure}[t] %% h=here, t=top,p=page of floats
\centerline{\resizebox{0.25\textheight}{!}{\rotatebox{0}
{\includegraphics{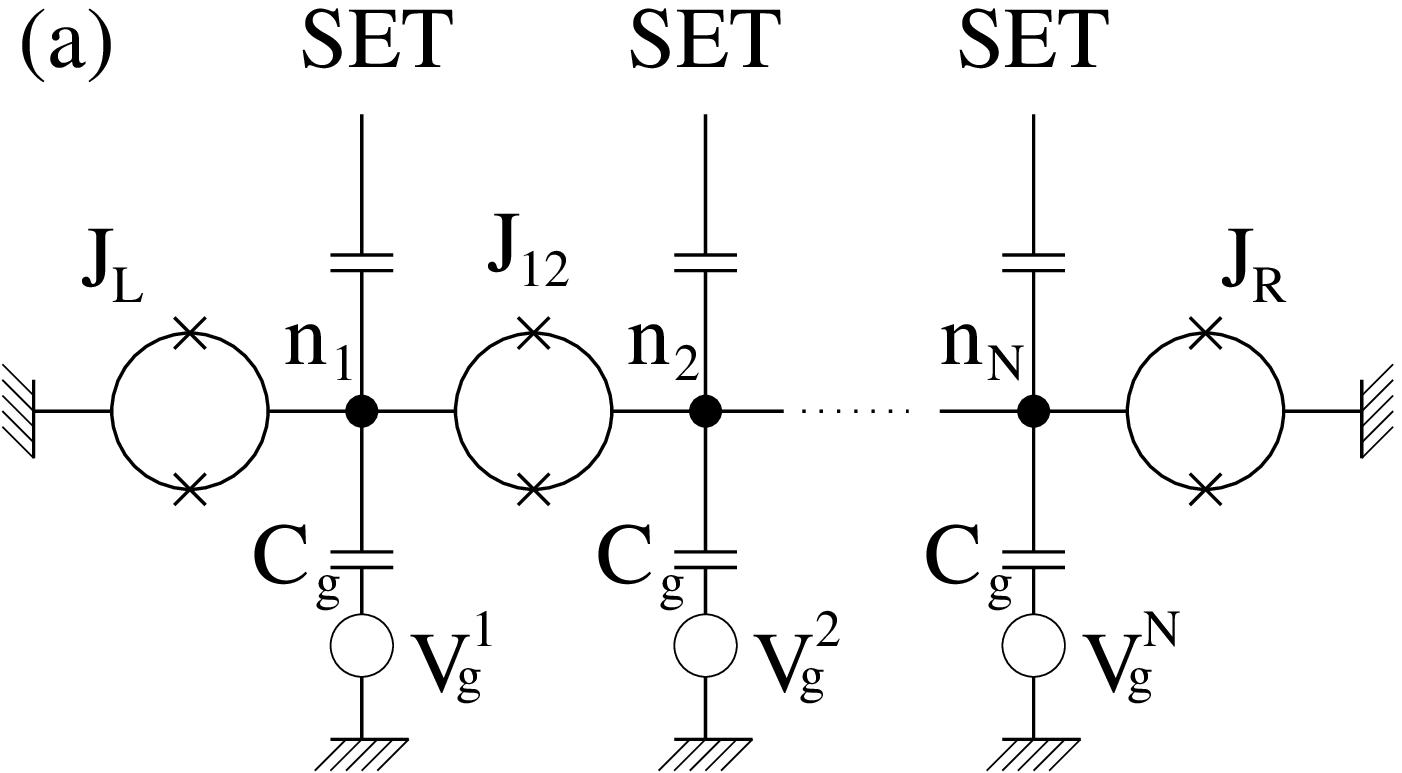}}}}
\centerline{\resizebox{0.28\textheight}{!}{\rotatebox{0}
{\includegraphics{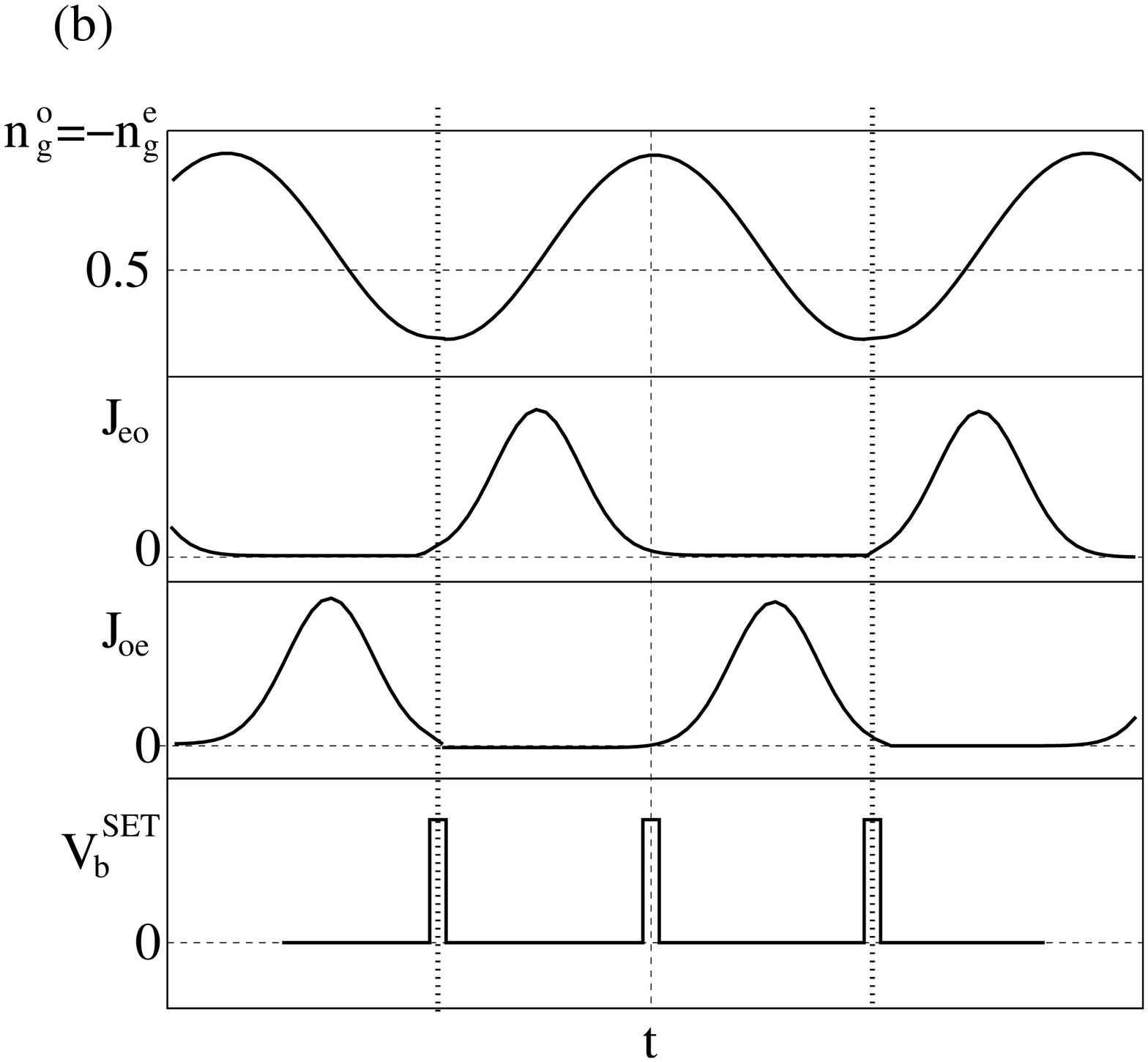}}}}
\caption{(a) The stabilized Cooper-pair pump. It is operated similarly
to the single-island system -- identical pulse sequence is applied to
every second junction (island) (b). Single electron transistors (SET) serve
as the dephasing elements. They are biased with short voltage pulses
every time the pump should be in a definite charge state [lowest plot
in (b)].}
\label{fi:setup}
\end{figure}

The system we consider is shown in Fig.~\ref{fi:setup}(a). It is
composed of $N$ Cooper-pair boxes. Nearest neighbours are coupled with
dc-SQUIDs, leftmost and rightmost islands are coupled to the
reservoirs via other SQUIDs (terminals). If the number of Cooper
pairs on the islands is $\{n_1, n_2, \ldots, n_N\}$, the electrostatic
energy of the system $E_{n_1, n_2, \ldots, n_N} = \sum_i E_{c(i)}(n_i
- n_{g (i)})^2 + E_m^{i, i+1}(n_i - n_{g (i)})(n_{i+1} - n_{g
  (i+1)})$. Here $E_{c(i)}$ are the charging energies of the islands
and $E_m^{i, i+1}$ is the energy of electrostatic coupling between two
neighboring islands, which is finite for finite capacitance of the
coupling SQUID. If the dimensionless gate charges, $n_{g(i)}= C_{g(i)}
V_{g(i)}/2e$ are close to $1/2$, the lowest energy states are
characterized by either zero or one Cooper pair on each island. With
this assumption we can again reduce the Hilbert space and map our
system to a finite anisotropic Heisenberg spin-$1/2$ chain in an
external magnetic field. The Hamiltonian
\begin{gather}
  H = -{1\over 2} B_x^1 \sigma _x^1 - {1\over 2}B_x^N \sigma _x^N -
  {1\over 2} \sum_{i=1}^{N} B_z^i \sigma _z^i
  \\
  + {1\over 2} \sum_{i=1}^{N-1} \left[\Delta_{i,i+1} \sigma _z^i
    \sigma _z^{i+1} \right.  \left. - J_{i, i+1} \left(\sigma _+^i
      \sigma _-^{i+1} + \sigma _+^{i+1} \sigma _-^{i}\right) \right] \nonumber
  \label{J42a}
\end{gather}
is characterized by constant electrostatic coupling amplitudes $\Delta
_{i,i+1}$ and tunable parameters -- $B_x^{1,N}$ - the Josephson
coupling of the leftmost and rightmost SQUIDs, $B_z^i$ - electrostatic
potential of the islands and $J_{i,i+1}$ - Josephson coupling between
neighbouring islands. In general the electrostatic coupling has a
finite range determined by the screening length $\lambda
=\sqrt{C_{J}/C_{g}}$ \cite{schoen}. As discussed below, we consider
the regime in which the electrostatic coupling is much weaker than the
on-site electrostatic energy. This implies that $\lambda \ll 1$ and we
can thus limit ourselves to nearest neighbour electrostatic coupling
only. 

Because of the electrostatic interaction, for $n_g^i$ close to $1/2$
the most favorable charge states are antiferromagnetic ($|010101\ldots
\rangle $ and $|101010\ldots \rangle $). The pumping procedure is a
simple generalization of the scheme described for one island and two
terminals. We start with one of the antiferromagnetic states and
transfer the charge of every island to the right by two sites to
achieve pumping. This can be implemented by applying identical pulse
sequence to every second island (and SQUID), {\it i.e.}  $B_x^1 =
J_{2,3} = J_{4,5} = \ldots \equiv J_{eo}$, $J_{1,2} = J_{3,4} = \ldots
= B_x^N \equiv J_{oe}$, $B_z^1 = B_z^3 = \ldots \equiv B_z^o$, and
$B_z^2 = B_z^4 = \ldots \equiv B_z^e$ [see Fig.~\ref{fi:setup}(b)]
($o,e$ denote odd and even islands respectively).  Here we focus on
systems with {\em odd} number of islands -- then the number of
coupling SQUIDs is even, and exactly half of them are active during
each step of the procedure. This symmetry simplifies our analysis but
is not crucial.

\begin{figure}[t] %% h=here, t=top,p=page of floats
  \centerline{\resizebox{0.28\textheight}{!}{\rotatebox{0}
      {\includegraphics{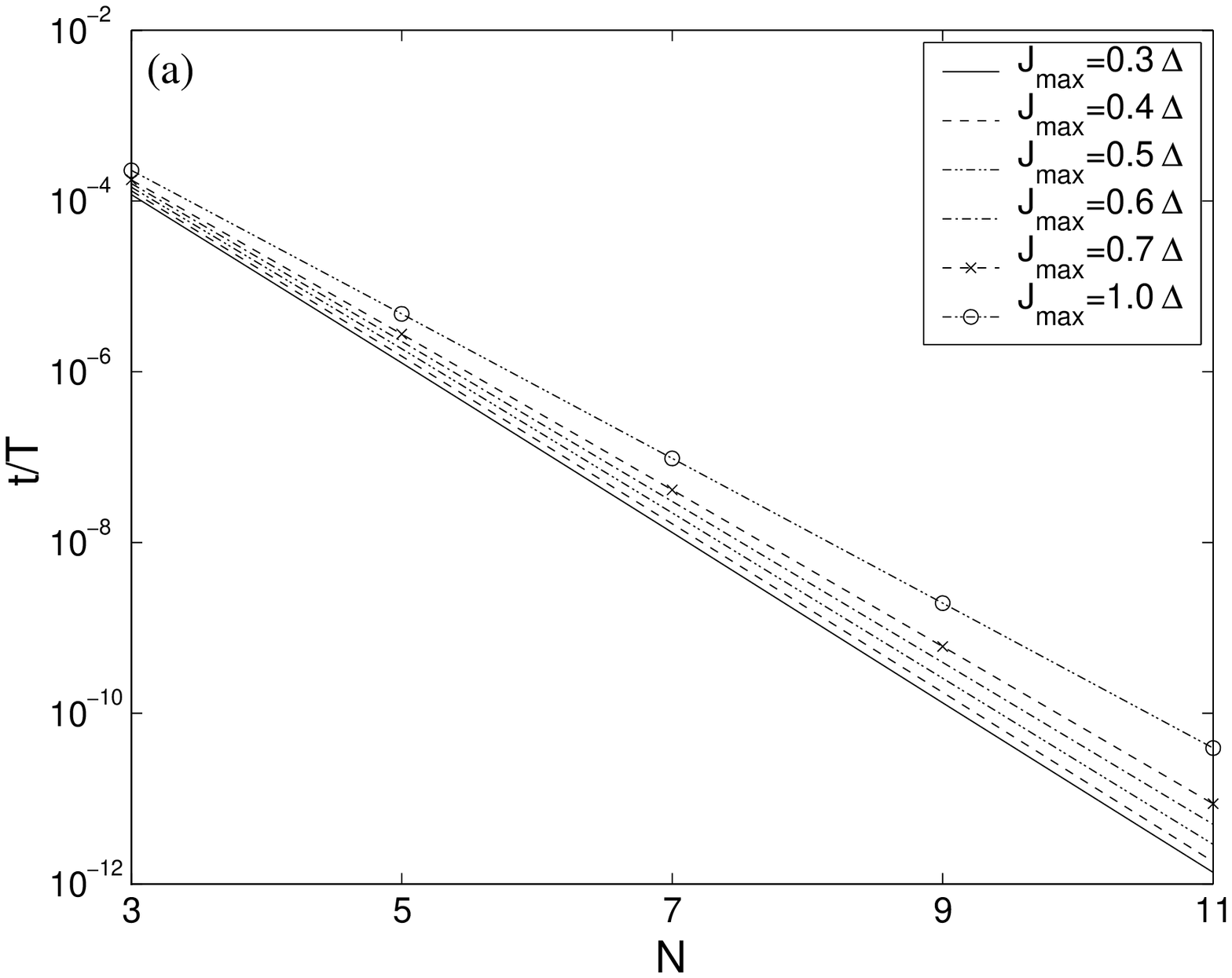}}}}
  \centerline{\resizebox{0.28\textheight}{!}{\rotatebox{0}
      {\includegraphics{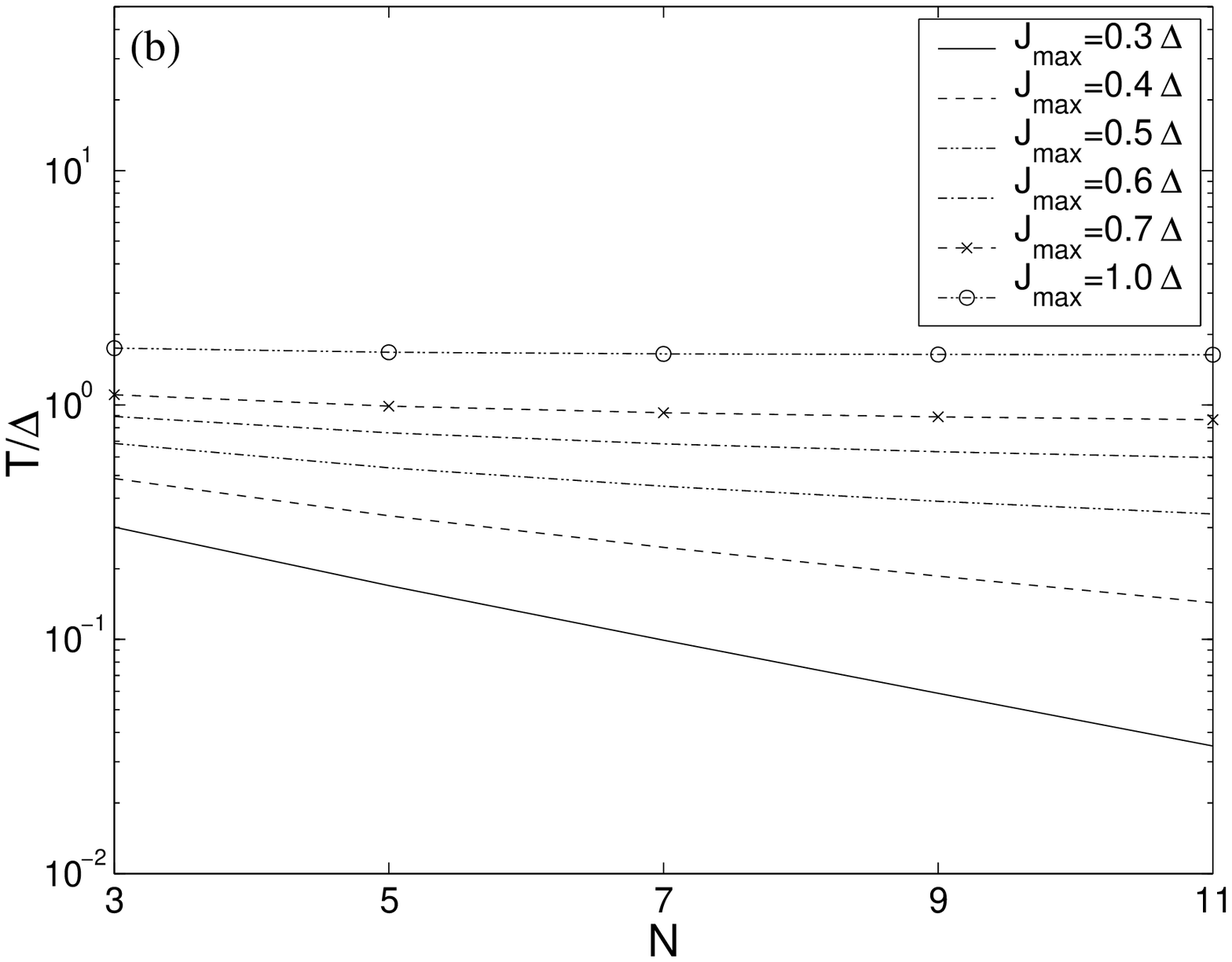}}}}
  \caption{(a) Pumping error $t/T$ as a function of $N$. For
    logarithmic vertical axis the dependence is linear, implying
    exponential suppression of the error with increasing $N$. (b)
    Transition rates between the charge states as a function of
    $N$.}
  \label{fi:results}
\end{figure}

The procedure, even for the single-island realization, takes place in
nondegenerate ground state. Due to the avoided degeneracy, spontaneous
transitions to unwanted levels cost energy and are thus suppressed.
The chain design additionally provides the following advantage: on the
one hand an energy gap between the antiferromagnetic states and the
remaining subspace $\Delta E \geq \min \{\Delta _{i,i+1}\}$, while on
the other hand accidental transitions between the antiferromagnetic
states need simultaneous transport through $(N+1)/2$ junctions --
another effect which diminishes exponentially with $N$ (see
Fig.~\ref{fi:results}(b) and explanation below). The most evident
advantage of this approach is, however, elimination of transport in
the wrong direction. Let us analyze the mechanism ruling this process.
Suppose that we start with the state $|101010 \ldots \rangle $. During
the first part of the cycle we want to shift the charge configuration
by one site to the right, {\it i.e.}  to arrive at $|010101\ldots
\rangle $. As mentioned before, the residual Josephson coupling of
closed SQUIDs leads to two ways of arriving at the same state. One
could anticipate interference effects here, but since the residual
coupling is a fluctuating quantity (controlled by fluctuating magnetic
field) we treat the process as incoherent. A simple analysis is based
on the assumption that a Cooper pair is transported through an open
(correct) SQUID with probability $p$, and through closed (wrong) SQUID
with the probability $q = 1-p$. The measure of this pumping error for
each island might be the ratio $q/p$. For a chain of $N$ junctions the
probability of correct shift becomes $\propto p^{N/2}$, and of
incorrect shift $\propto q^{N/2}$.  Hence the pumping error
$(q/p)^{N/2}$ decreases {\em exponentially} with the number of
junctions in the chain. More rigorous analysis requires calculation of
the energy difference between the two lowest energy levels for $B_z^i
= 0$. Indeed, in this case the lowest energy eigenstates are symmetric
and antisymmetric superpositions of the antiferromagnetic states (with
some addition of other charge states), and the energy difference,
being the rate of phase progression, in the charge basis is equivalent
to the rate of transition.  Fig.~\ref{fi:results}(a) shows the results
for various values of the maximum Josephson coupling of the SQUIDs.
$T$ is the rate calculated for the desired transitions only (assuming
no residual coupling), $t$ is calculated for the wrong path.  Then,
the time of single operations should be much larger than $1/T$ for the
adiabatic condition to be satisfied.  The numerical calculation is
performed for a realistically achievable residual coupling equal to
$1\%$ of the maximum Josephson coupling of the SQUIDs. The
values of the electrostatic coupling amplitudes, and maximum Josephson
couplings are for simplicity assumed  identical for each
junction (this  is, again, not crucial and our observations
are valid also for different settings).  The pumping error vs. $N$ is
plotted in Fig.~\ref{fi:results}(a).
The dependence is clearly exponential, and very weakly depends on the
value of the maximum Josephson coupling. However, the value of $T$
itself should not decrease too strongly with increasing $N$ as it
determines the adiabatic rates. From Fig.~\ref{fi:results}(b) we can
see that the ratio $J_{max}/ \Delta $ should be as close to $1$ as
possible to avoid the effects of exponential suppression. For this
case, $T/ \Delta \approx 1$, where $\Delta $ can be chosen large thus
yielding a very comfortable adiabatic condition.  Note that even for
$J_{max}/ \Delta \approx 1$ we can still work in the reduced Hilbert
space, as long as the charging energies of individual islands are much
larger than $J_{max}$ (this means that the system should be operated
in the charging regime with weak electrostatic coupling). The
exponential suppression of the energy gap between the lowest states
implies that unwanted transitions between the antiferromagnetic states
also diminish very rapidly with increasing $N$ (note that this
suppression for very small, erroneous, values of Josephson coupling is
much stronger than for moderate values, expected for open terminals).
This of course does not mean that the optimal setup should be
represented by as long an array as possible. Most probably there is
some moderate number of junctions for which the pumping errors are
strongly suppressed, and the decoherence effects typical for
macroscopic system are yet not severe. However, examination of the
long-chain limit goes beyond our analysis and would be appropriate for
a separate publication.

Although during the passage the state of the system is a superposition
of numerous charge configurations, when all Josephson couplings are
switched off (or rather intended to be) the pump should be in one of
the antiferromagnetic states. All other components (nonadiabatic
corrections, residual Josephson coupling effects) contribute to
pumping errors. The optimal solution would be to project the state
onto a charge state by performing charge measurement. Indeed, if
the measured state is very close to an eigenstate of the measured
observable, the measurement takes the state closer to the
eigenstate. If frequently repeated, the measurement can diminish the
probability of transition to other eigenstates arbitrarily. This
phenomenon, known as quantum Zeno effect \cite{misra}, can be utilized
also in our system. The charge measurement can be performed by a
single-electron transistor (SET) coupled capacitively to the
superconducting island \cite{makhlin2} [see
Fig.~\ref{fi:setup}(a)]. This measurement is characterized by
few time scales, the dephasing time $\tau _{\phi }$, during which the
off-diagonal density matrix  elements (in the basis of the
measured observable) vanish, the measurement time, $\tau _{meas}$,
after which the information about the charge can be read out, and the
mixing time, after which the back action of the SET on the island
destroys the information about the state. The parameters can be
selected in such way that $\tau _{mix} \geq \tau _{meas} \geq \tau
_{\phi }$. Since the actual information about the state is not
important, and only the dephasing mechanism modifies the state, the
quantum Zeno projection can be realized within a relatively short
time. Moreover, even if the time available is too short for the
off-diagonal entries of the density matrix to vanish, partial
dephasing still brings the system closer to the desired charge
state. The system can be thus dephased each time the system is {\em
supposed to be} in a definite charge state by switching on the bias
voltage of the SETs for a very short time interval.

In summary, the presented scheme provides an accurate pumping
procedure. The array design exponentially suppresses the most severe,
untracable error of reversed current flow and diminishes errors due to
undesired charge configurations. These errors can be further minimized
by repeated dephasing with single-electron transistors.

The authors thank J.~Siewert, O.~Buisson, and S.~Paraoanu for
discussions. M.C. acknowledges support from Polish Ministry Grant No.
1 P03B 067 30.

\end{document}